\documentclass[aps,prx,reprint,superscriptaddress,nofootinbib,longbibliography]{revtex4-2}

\usepackage[utf8]{inputenc}
\usepackage[T1]{fontenc}
\usepackage{lmodern}
\usepackage{microtype}
\usepackage{mathtools}

\usepackage{placeins}

\usepackage{amsmath}
\usepackage{amsfonts}
\usepackage{graphicx}
\usepackage{hyperref}
\hypersetup{
  colorlinks=true,
  citecolor=blue,
  linkcolor=blue,
  urlcolor=blue
}

\usepackage{mdframed}
\usepackage{comment}
\usepackage[braket,qm]{qcircuit}
\usepackage{multirow}

\begin{document}

\title{Categorizing Readout Error Correlations on Near Term Quantum Computers}

\author{Benjamin Nachman}
\email{bpnachman@lbl.gov}
\affiliation{Physics Division, Lawrence Berkeley National Laboratory, Berkeley, CA 94720, USA}

\author{Michael R. Geller}
\affiliation{Center for Simulational Physics, University of Georgia, Athens, Georgia 30602, USA}

\begin{abstract}
Readout errors are a significant source of noise for near term quantum computers.  A variety of methods have been proposed to mitigate these errors using classical post processing.  For a system with $n$ qubits, the entire readout error profile is specified by a $2^n\times 2^n$ matrix.  Recent proposals to use sub-exponential approximations rely on small and/or short-ranged error correlations.  In this paper, we introduce and demonstrate a methodology to categorize and quantify multiqubit readout error correlations. Two distinct types of error correlations are considered: sensitivity of the measurement of a given qubit to the state of nearby ``spectator'' qubits, and measurement operator covariances. We deploy this methodology on IBMQ quantum computers, finding that error correlations are indeed small compared to the single-qubit readout errors on IBMQ Melbourne (15 qubits) and IBMQ Manhattan (65 qubits), but that correlations on IBMQ Melbourne are long-ranged and do not decay with inter-qubit distance.
\end{abstract}

\date{\today}
\maketitle

\section{Introduction}

Quantum computers have the potential to provide an exponential speedup for a variety of scientific and industrial applications.  However, the present noisy intermediate-scale quantum (NISQ)~\cite{Preskill2018quantumcomputingin} computers introduce significant errors that must be mitigated to acquire useful output.  These errors can be categorized into three types: state preparation errors, gate errors, and measurement errors.  State preparation errors occur when a quantum computer is initialized to the ground state, but some qubits are in an excited state.  Gate errors occur when unitary operations are applied to the computer, leaving the qubits in a superposition of the desire state and noisy states.  Measurement errors can often dominate state-preparation and gate errors, partly because the readout time is not much shorter than the decoherence time, so that decoherence can occur during the measurement process. We note that in most practical applications, it is not possible to disentangle state preparation and measurement (SPAM) errors, so these will be considered together in this work.


Error correction on a quantum computer is significantly different than on classical computers because qubits cannot be copied (`cloned')~\cite{Park1970,Wootters:1982zz,DIEKS1982271}.  The analog of copying is embedding a logical qubit into an entangled state of several physical qubits.  Full quantum error correction using an entangling strategy requires significant overhead in terms of additional qubits and quantum gates.  Therefore, fault tolerance is infeasible on NISQ hardware due to the limited size and fidelity of near term qubit arrays.  Instead, error mitigation (and not correction) is the general approach to quantum errors in the NISQ era.

A variety of methods have been proposed to mitigate SPAM errors~\cite{bialczak_quantum_2010,neeley_generation_2010,dewes_characterization_2012,magesan_machine_2015,debnath_demonstration_2016,song_10-qubit_2017,gong_genuine_2019,wei_verifying_2020,havlicek_supervised_2019,chen_detector_2019,chen_demonstration_2019,maciejewski_mitigation_2020,urbanek_quantum_2020,nachman_unfolding_2020,hamilton_error-mitigated_2019,200601805,karalekas_quantum-classical_2020,181010523,geller_rigorous_2020,2010.07496,Funcke:2020olv,201008520,210102331}.  The fundamental object in these schemes is the \textit{response matrix}, $R=\Pr(\text{measure state $i$}|\text{state is $j$})$.  If a quantum computer has $n$ qubits, then $R\in[0,1]^{n\times n}$.  Constructing the full response matrix requires an exponential number of measurements: each of the $2^n$ states is prepared and measured many times to quantify the migrations away from the prepared state during the readout process.  Many of the schemes referenced above propose approximations to $R$ that only require polynomial measurements.  For example, one possibility is to assume that qubit measurement errors are independent from each other \cite{gong_genuine_2019}.  Then, a user only needs to prepare $2n$ states to measure $\Pr(\text{measure 0}|\text{state was 1})$ and $\Pr(\text{measure 1}|\text{state was 0})$ for each qubit. In this work, we will explore multiqubit readout errors in order to understand to what extent schemes that ignore correlations are applicable to existing devices. Previous work has reported multiqubit correlations \cite{181010523,200601805,201008520,210102331}, but their size and range have not been systematically investigated.

Our paper is organized as follows.  Section~\ref{sec:method} introduces our protocol for quantifying SPAM errors.  Experimental results are presented in Sec.~\ref{sec:results}, and the paper ends with our conclusions and outlook in Sec.~\ref{sec:conclusions}.

\section{Methods}
\label{sec:method}

Let $E_j^{(i)}$ be the positive-operator-valued measure (POVM) element $j$ for qubit $i$ (in the ideal projective limit, $E_i=\ket{i}\bra{i}$).  We define symmetrized, per qubit readout errors $\epsilon$ as

\begin{align}
    \epsilon_i = \frac{1}{2}\left(\langle E_1^i\rangle_{\ket{0\cdots0}}+\langle E_0^i\rangle_{\textbf{X}_i\ket{0\cdots 0}}\right)\,,
\end{align}
where $\langle \bullet\rangle_\psi$ represents the expectation value of operator $\bullet$ for the state $\psi$ and $\textbf{X}_i$ is the NOT operator acting on qubit $i$ (i.e. the Pauli operator $\sigma_j^x$).  Our goal is to investigate readout error correlations and compare their magnitude to $\epsilon$.  We investigate two families of $2$-qubit readout error correlations.  The first, which we call $A$-type correlators, are defined as
\begin{align}
    A_{ij}=\langle E_1^i\rangle_{\ket{0\cdots 0}}-\langle E_1^i\rangle_{{\bf X}_j\ket{0\cdots 0}}\,.
\label{defA}
\end{align}
This quantifies the dependence of the qubit $i$  readout error on the state of a nearby ``spectator'' qubit $j$.
The second family, which we call $C$-type correlators, are measurement operator covariances
\begin{align}
    C_{ij}=\langle E_0^iE_0^j\rangle_{\psi}-\langle E_0^i\rangle_{\psi}\langle E_0^j\rangle_{\psi}\,,
\end{align}
where $\psi$ is any state.  In this paper, we use $\psi=\ket{0}^{\otimes n}$.  Computing $\epsilon,A$, and $C$ requires the preparation of $n+1$ states: the ground state and then the $n$ states where each qubit is in the $1$ state and every other qubit is in the zero state.  These states are measured up to 820k times to reduce the statistical noise. These correlators were measured previously on Melbourne \cite{181010523}. (However, our definition of $A_{ij}$ differs from that of \cite{181010523} by a sign.)


\begin{figure}[h!]
    \centering
    \includegraphics[width=0.4\textwidth]{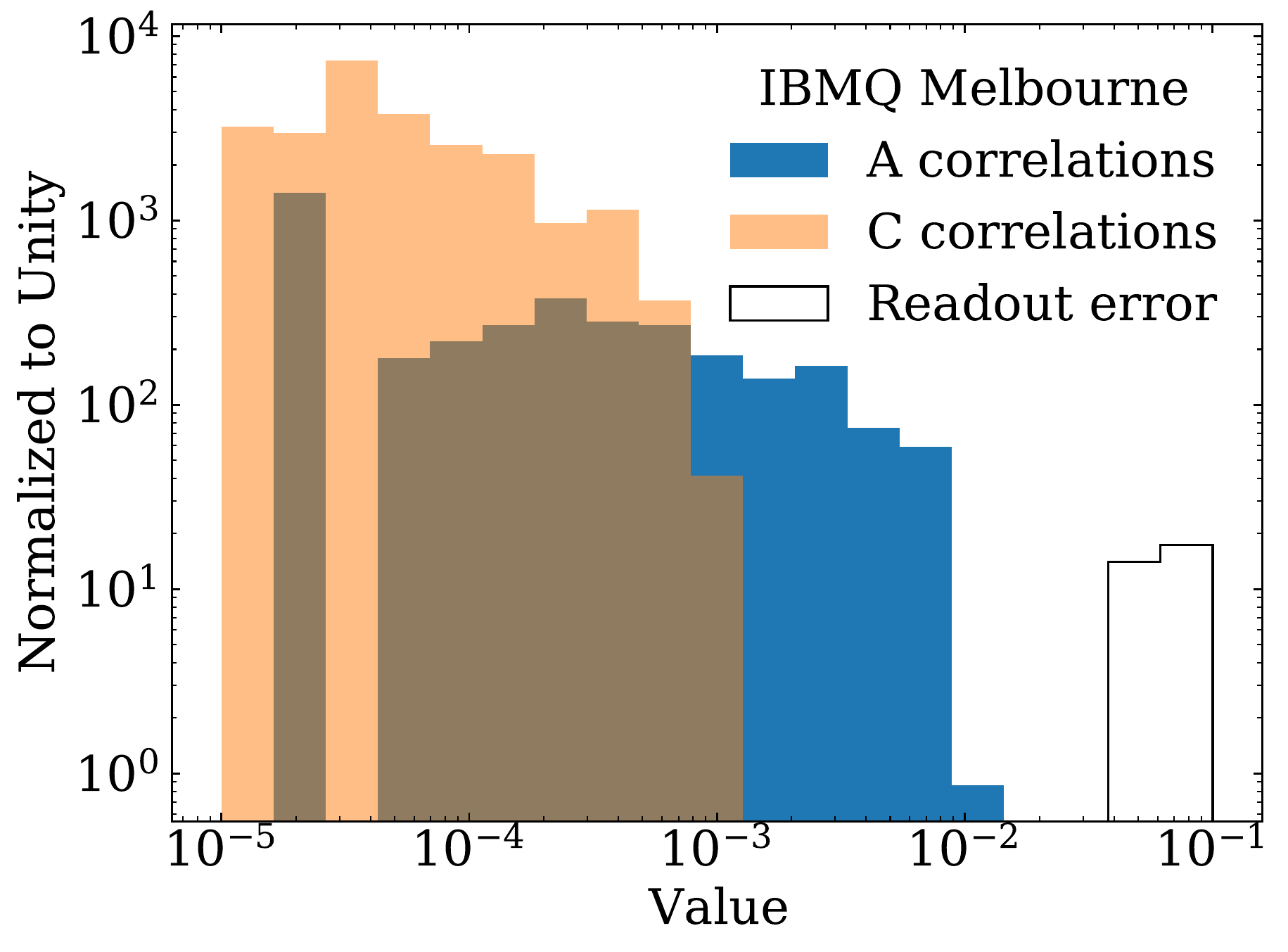}
    \includegraphics[width=0.4\textwidth]{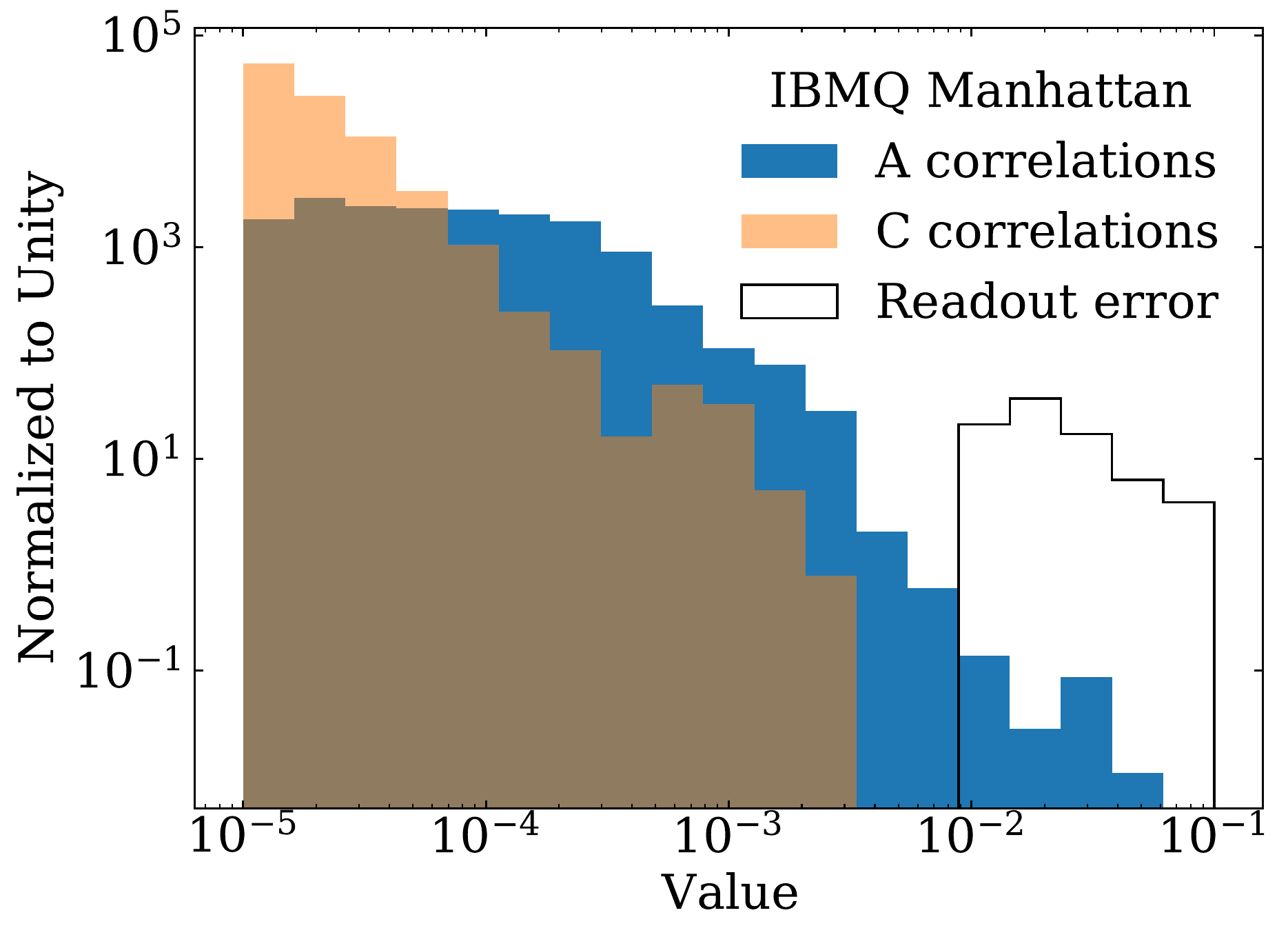}
    \caption{Histograms of the $\epsilon,A$, and $C$ values for the IBMQ Melbourne computer (top) and the IBMQ Manhattan computer (bottom).  Note that the first bin does not include underflow.}
    \label{fig:aij_manhattan}
\end{figure}

\section{Results}
\label{sec:results}

We implement all of the circuit preparation and scheduling with IBM's~\texttt{Qiskit}~software~\cite{Qiskit} and use the IBMQ Melbourne computer (all 15 qubits) and the IBMQ Manhattan computer (all 65 qubits).  We use 81920 (819200) measurements for each circuit on IBMQ Melbourne (Manhattan).

\begin{figure}[h!]
    \centering
    \includegraphics[width=0.4\textwidth]{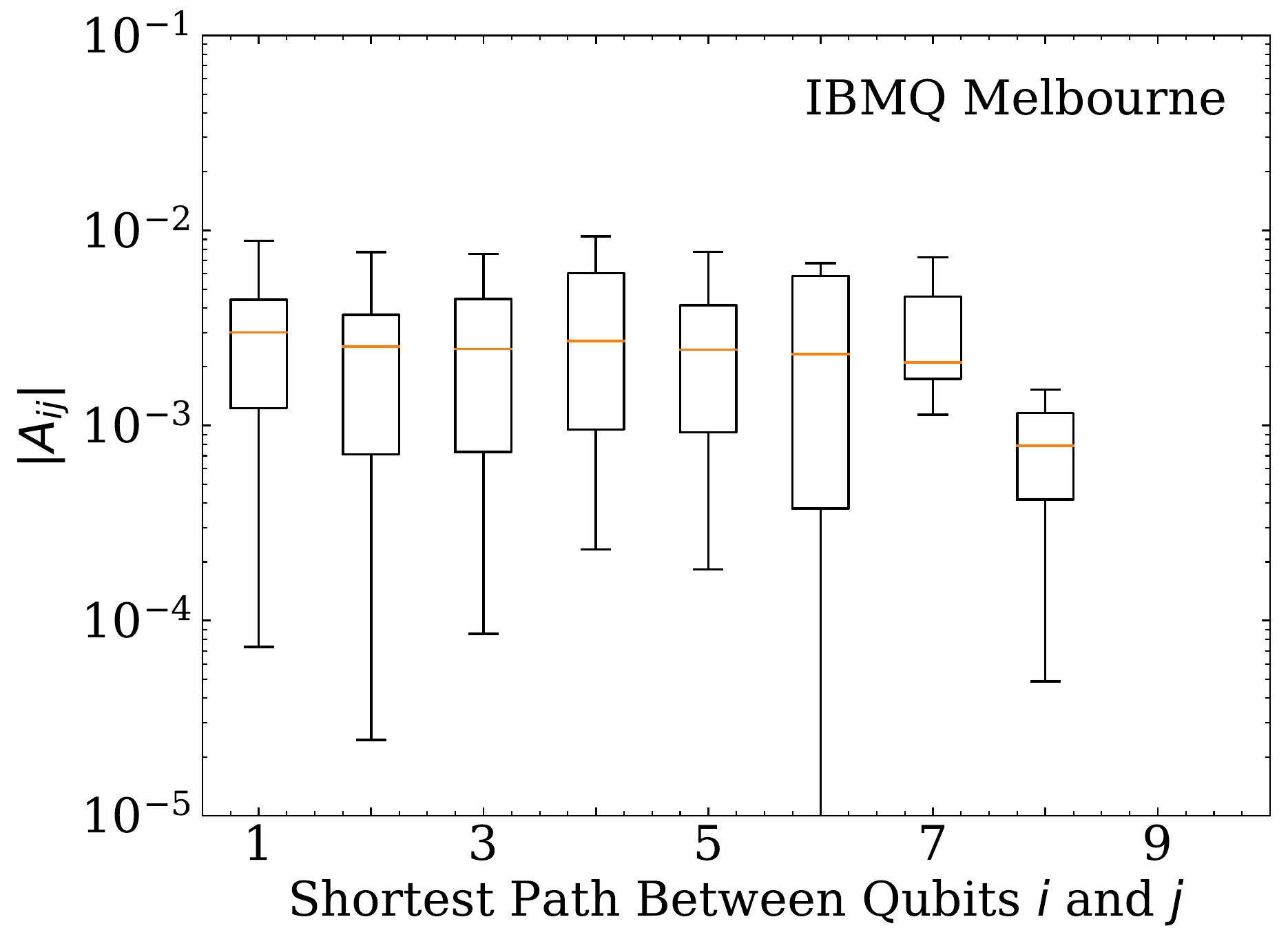}
    \includegraphics[width=0.4\textwidth]{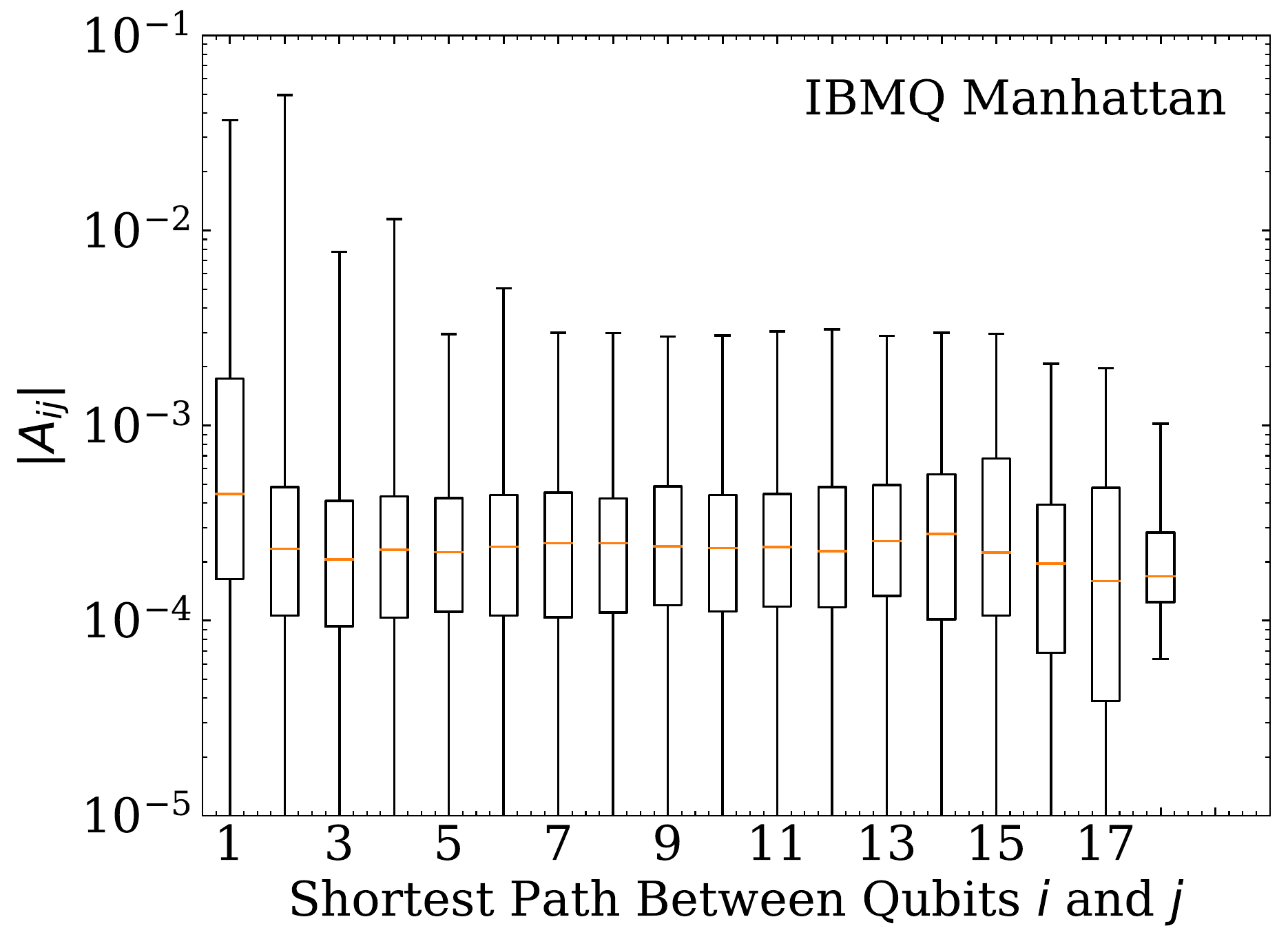}
    \caption{The minimum, first quartile, median, third quartile, and maximum $|A_{ij}|$ in bins of the shortest connected path between qubits $i$ and $j$ for IBMQ Melbourne (top) and IBMQ Manhattan (bottom).  }
    \label{fig:aij_distance}
\end{figure}

Histograms of the $\epsilon,A$, and $C$ values for these two computers are presented in Fig.~\ref{fig:aij_manhattan}.  The per-qubit readout errors $\epsilon_i$ range from 1\% to 10\% in both machines, with systematically lower readout errors on IBMQ Manhattan.  The $A$ correlations are typically higher than the $C$ correlations, with a small fraction of $A$ errors reaching the 1-10\% level on IBMQ Manhattan.

To examine the spatial distribution of the dominant ($A$-type) correlations across the devices, we correlate each measured $A_{ij}$ with the minimum connected distance between qubits.  We define this distance by computing the number of qubits coupled by two-qubit gates on the IBMQ machines that are required to connect two given qubits. In other words, we measure inter-qubit distance by the minimum number of CNOT-active edges connecting qubits $i$ and $j$. This distance is computed efficiencly using Dijkstra's algorithm~\cite{Dijkstra.1959}.  Figure~\ref{fig:aij_distance} presents the average $A_{ij}$ in bins specifying the minimum connected distance.  Interestingly, there is nearly no correlation between the size of the $A_{ij}$ and qubit-qubit distance for IBMQ Melbourne, while on IBMQ Manhattan, the noisiest pair of qubits are also the ones that have the shortest path between them. We interpret these data as showing that the measurement error correlations on Melbourne, while typically smaller than \%1 in magnitude, are long ranged, extending over the entire device.  On Manhattan the error correlations are larger but are short-ranged.

Finally, we discuss the role of sampling errors in our experiment. Figure~\ref{fig:aij_manhattan} shows that the two-qubit correlators $A$ and $C$  take a wide range of values on Melbourne and Manhattan. At the same time, the $A$ and $C$ values themselves have statistical errors resulting from their estimation with a finite number of measurement samples. The sampling errors therefore bound the size of the smallest $A$ and $C$ values that are statistically meaningful. Estimating the output probability $p_x$ for a single classical state $x \in \{0,1\}^n$ by using $N$ measurement samples results in a standard error for the noisy estimate $p_x^{\rm est}$ given by
\begin{equation}
\sqrt{{\rm Var}(p_x^{\rm est})} = 
\sqrt{\frac{ p_x(1-p_x)}{N}},
\end{equation}
which is no larger than $(2\sqrt{N})^{-1}$. 
Each $\epsilon_i$ is measured by combining two estimated probabilities, so its standard error is bounded by $(\sqrt{2N})^{-1}$. Each $A_{ij}$ is measured by subtracting two estimated probabilities, so its standard error is also upper bounded by $(\sqrt{2N})^{-1}$. Each $C_{ij}$ is obtained from a two-qubit probability distribution, so its standard error is bounded by $(2\sqrt{N})^{-1}$. Therefore the bound $(\sqrt{2N})^{-1}$, which takes the value of $2.5 \times 10^{-3}$ on Melbourne ($N \! = \! 81920$) and $7.8 \times 10^{-4}$ on Manhattan ($N \! = \! 819200$), can be applied to all three measured quantities. This means that the long-range correlations observed on Melbourne are somewhat above the noise floor, whereas the long-range correlations on Manhattan are consistent with statistical noise.

\section{Conclusions}
\label{sec:conclusions}

In this paper, we present a set of observables that can be used for categorizing and quantifying the correlated measurement errors on near-term quantum computers. In addition to single-qubit SPAM errors $\epsilon_i$, we measured a family of two-qubit correlators $A_{ij}$ that quantify sensitivity to spectator qubits, as well as POVM covariances $C_{ij}$. We find that the $C$-type correlations are always subdominant to the per-qubit readout errors on IBMQ Melbourne and Manhattan, while the $A$-type correlations can be as large as the $\epsilon_i$ on IBMQ Manhattan, reaching several percent. Furthermore, the $A$-type correlations are long-ranged on Melbourne and apparently extend over the entire device, whereas on Manhattan they are short-ranged and confined to neighboring qubits.  This is good news for future scalable measurement error mitigation as well as fault-tolerant quantum computation, which both require error correlations to be bounded in space and time.

\section*{Code and Data}

The code for this paper can be found at \url{https://github.com/LBNL-HEP-QIS/ReadoutErrors}.  Quantum computer data are available upon request.

\begin{acknowledgments}

BN is grateful to C. Bauer, M. Freytsis, W. de Jong, and M. Urbanek for useful comments on the manuscript.  This work is supported by the U.S. Department of Energy, Office of Science under contract DE-AC02-05CH11231. In particular, support comes from Quantum Information Science Enabled Discovery (QuantISED) for High Energy Physics (KA2401032) and the Office of Advanced Scientific Computing Research (ASCR) through the Accelerated Research for Quantum Computing Program.   This research used resources of the Oak Ridge Leadership Computing Facility, which is a DOE Office of Science User Facility supported under Contract DE-AC05-00OR22725.

\end{acknowledgments}

\bibliography{myrefs}

\appendix

\section{Full Matrix Information}
\label{sec:full}

\begin{figure}[h!]
    \centering
    \includegraphics[width=0.5\textwidth]{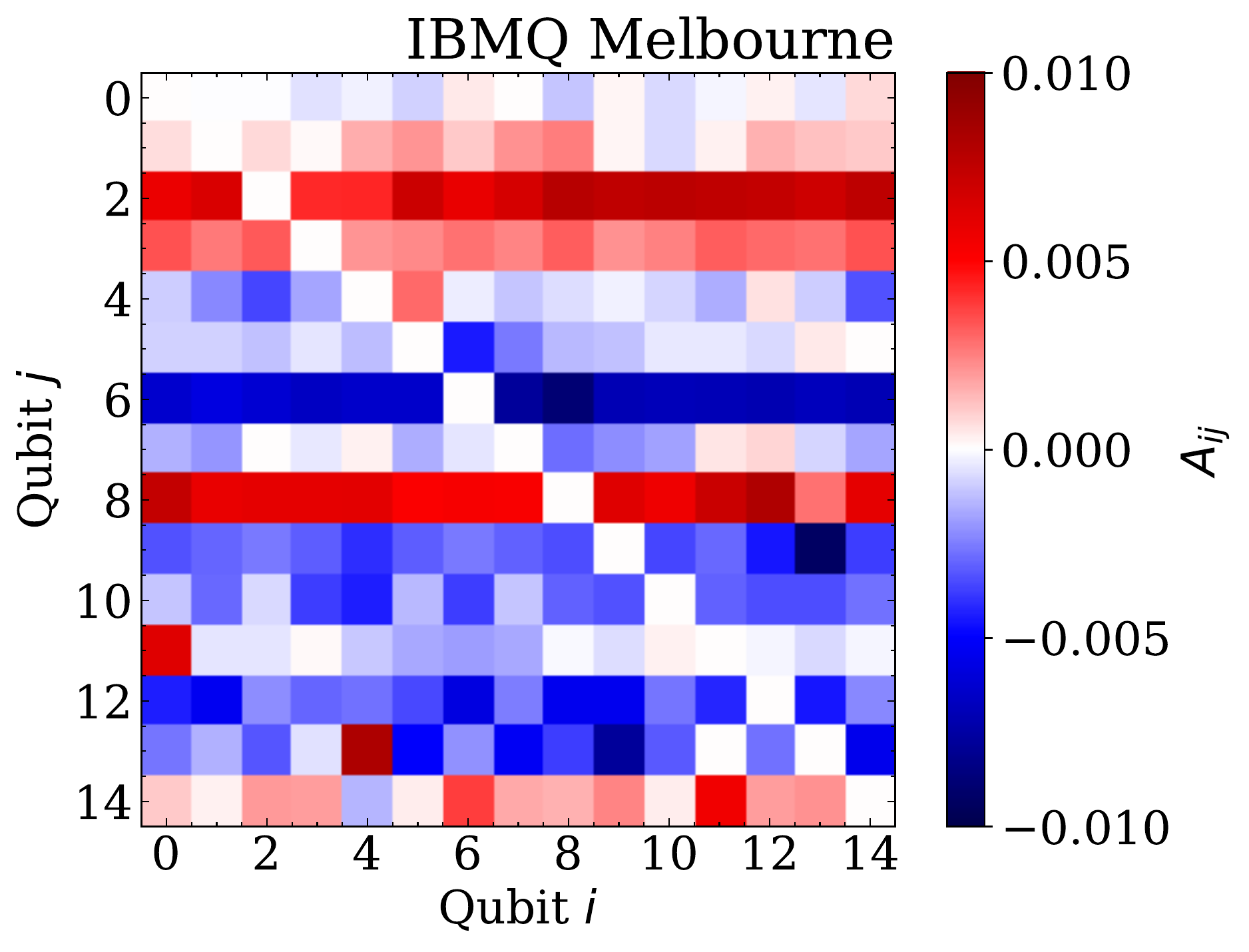}
    \includegraphics[width=0.5\textwidth]{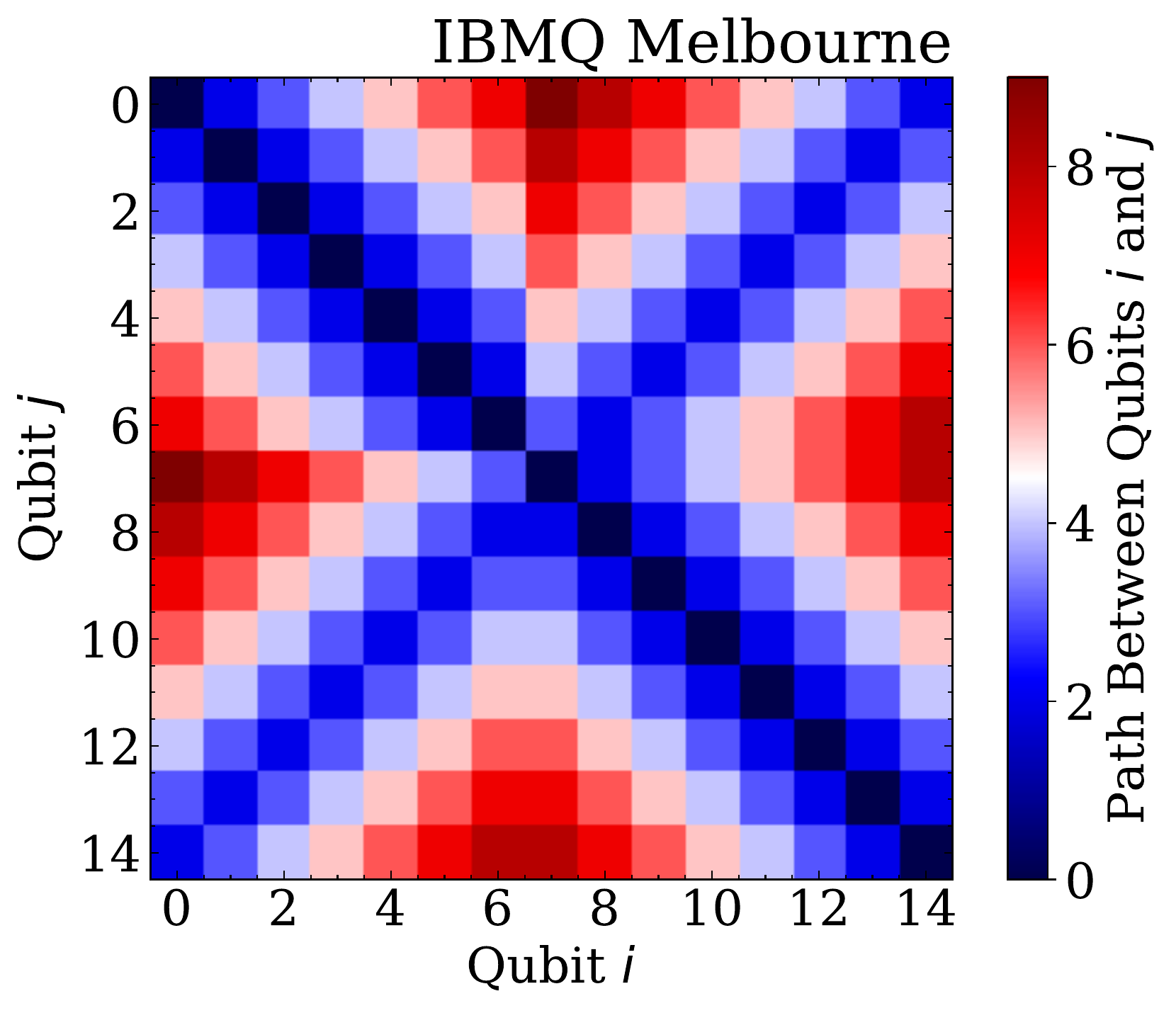}
    \caption{The matrix $A_{ij}$ (top) and the minimum distance between qubits (bottom) for the IBMQ Melbourne computer.}
    \label{fig:aijdij_melbourne}
\end{figure}

\begin{figure}[h!]
    \centering
    \includegraphics[width=0.5\textwidth]{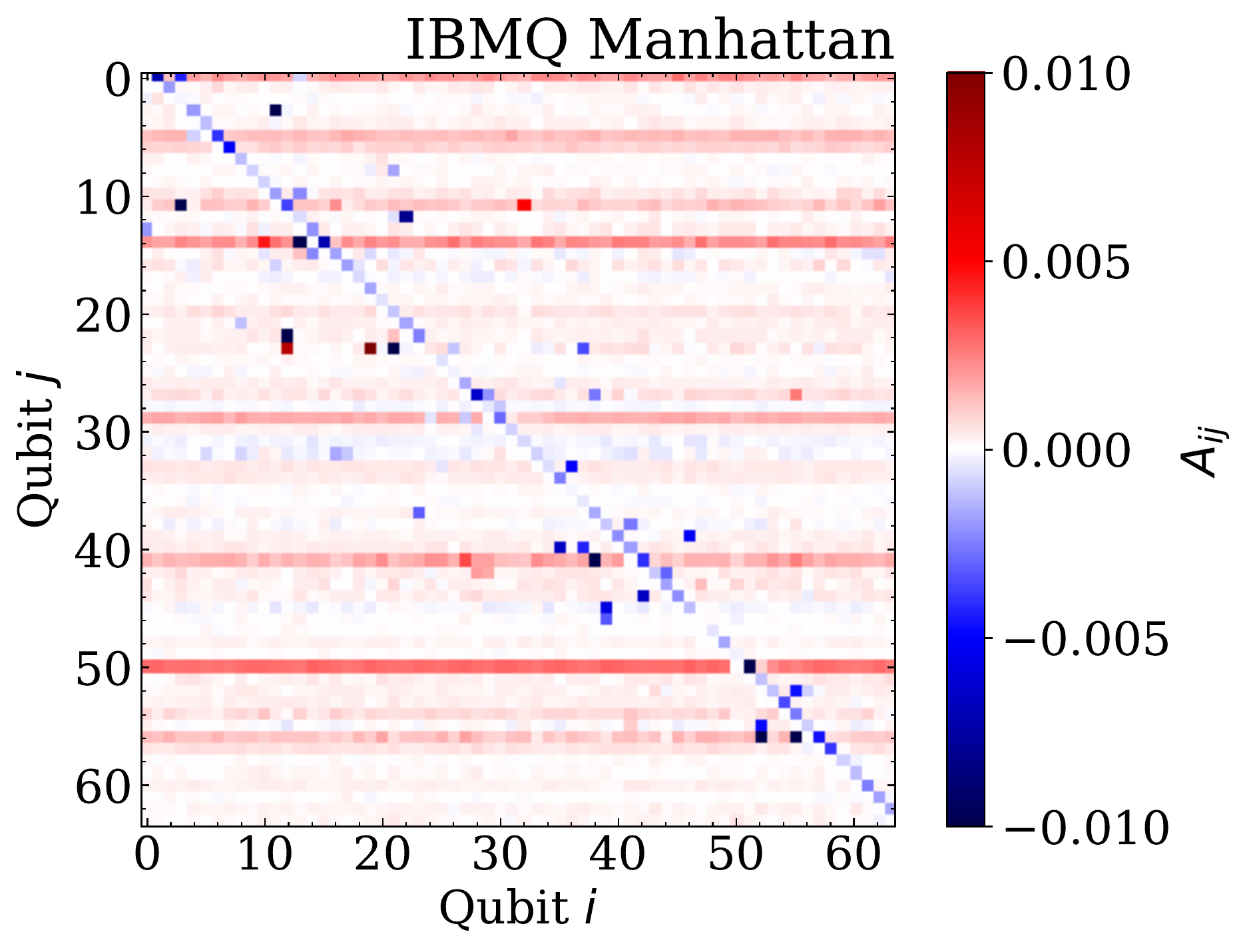}
    \includegraphics[width=0.5\textwidth]{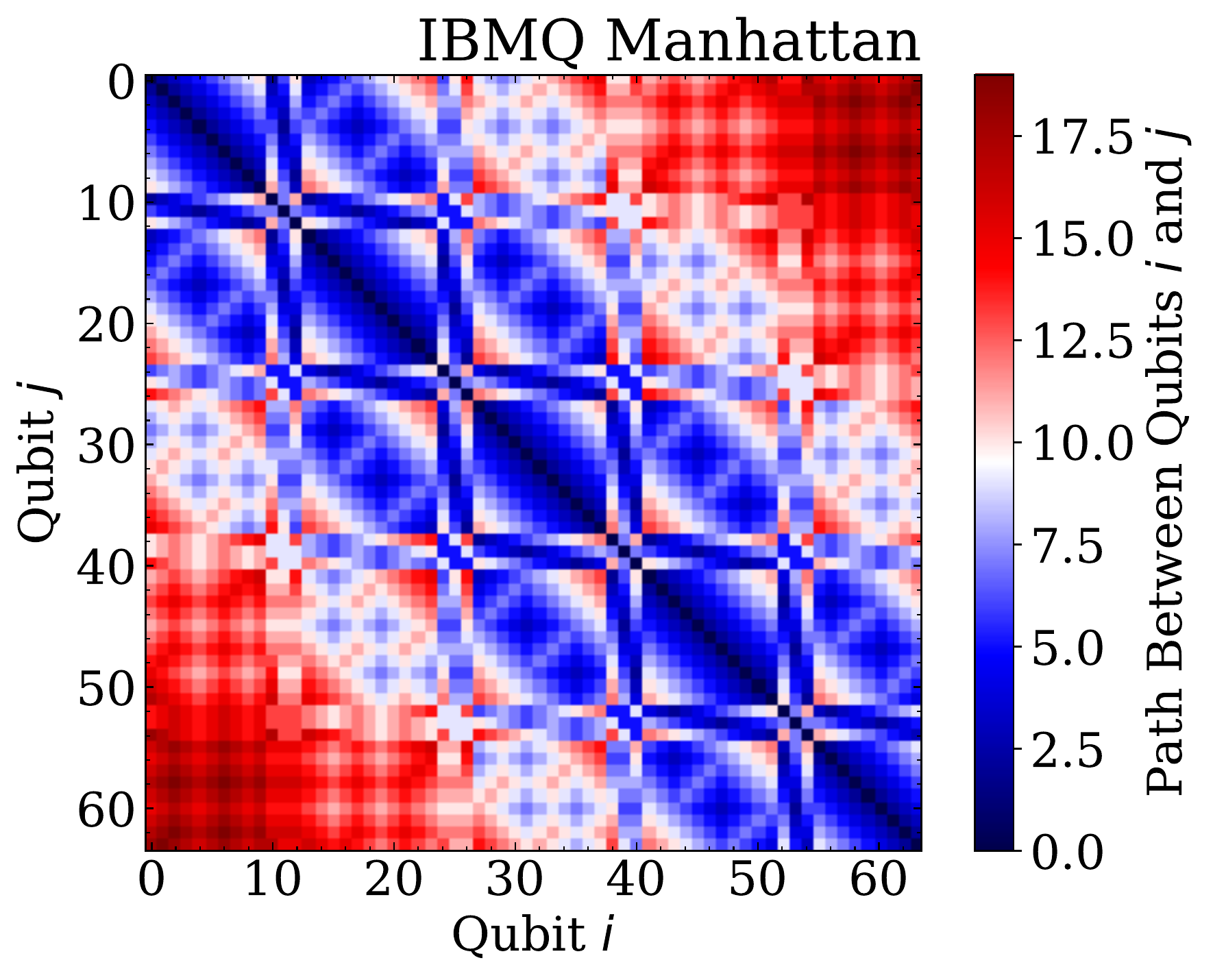}
    \caption{The matrix $A_{ij}$ (top) and the minimum distance between qubits (bottom) for the IBMQ Manhattan computer.}
    \label{fig:aijdij_manhattan}
\end{figure}

The full $A_{ij}$ and minimum distance matrices for IBMQ Melbourne and Manhattan are presented in Fig.~\ref{fig:aijdij_melbourne} and~\ref{fig:aijdij_manhattan}, respectively.  The correlation between nearness and high errors in $A_{ij}$ for Manhattan from Fig.~\ref{fig:aij_distance} can be observed as a slightly off-diagonal blue stripe in Fig.~\ref{fig:aijdij_manhattan}.

\end{document}